\documentclass[twoside,twocolumn,9pt]{article}
\usepackage{gensymb}
\usepackage{extsizes}
\usepackage[numbers,super,sort&compress,comma]{natbib}
\usepackage[version=3]{mhchem}
\usepackage[left=1.5cm, right=1.5cm, top=1.785cm, bottom=2.0cm]{geometry}
\usepackage{balance}
\usepackage{mathptmx}
\usepackage{sectsty}
\usepackage{graphicx}
\usepackage{lastpage}
\usepackage[format=plain,justification=justified,singlelinecheck=false,font={stretch=1.125,small,sf},labelfont=bf,labelsep=space]{caption}
\usepackage{float}
\usepackage{fancyhdr}
\usepackage{fnpos}
\usepackage[english]{babel}
\addto{\captionsenglish}{%
  
}
\usepackage{array}
\usepackage{droidsans}
\usepackage{charter}
\usepackage[T1]{fontenc}
\usepackage[usenames,dvipsnames]{xcolor}
\usepackage{setspace}
\usepackage[compact]{titlesec}
\usepackage{hyperref}
\usepackage{amsmath}
\usepackage{booktabs}
\usepackage{siunitx}
\DeclareSIUnit\angstrom{\text{Å}}

\usepackage{epstopdf}

\definecolor{cream}{RGB}{222,217,201}

\csname @ifundefined\endcsname{endmcitethebibliography}
  {}{}
\providecommand*{\mciteSetBstSublistMode}[1]{}
\providecommand*{\mciteSetBstMaxWidthForm}[2]{}

\providecommand*{\mciteBstWouldAddEndPunctfalse}{\let\EndOfBibitem\relax}
\providecommand*{\mciteSetBstMidEndSepPunct}[3]{}
\providecommand*{\mciteSetBstSublistLabelBeginEnd}[3]{}
\providecommand*{\EndOfBibitem}{}

\begin{document}

\pagestyle{fancy}
\thispagestyle{plain}
\fancypagestyle{plain}{
\renewcommand{\headrulewidth}{0pt}
}

\makeFNbottom
\makeatletter
\renewcommand\LARGE{\@setfontsize\LARGE{15pt}{17}}
\renewcommand\Large{\@setfontsize\Large{12pt}{14}}
\renewcommand\large{\@setfontsize\large{10pt}{12}}
\renewcommand\footnotesize{\@setfontsize\footnotesize{7pt}{10}}
\makeatother

\renewcommand{\thefootnote}{\fnsymbol{footnote}}
\renewcommand\footnoterule{\vspace*{1pt}%
\color{cream}\hrule width 3.5in height 0.4pt \color{black}\vspace*{5pt}}
\setcounter{secnumdepth}{5}

\makeatletter
\renewcommand\@biblabel[1]{#1}
\renewcommand\@makefntext[1]%
{\noindent\makebox[0pt][r]{\@thefnmark\,}#1}
\makeatother
\renewcommand{\figurename}{\small{Fig.}~}
\sectionfont{\sffamily\Large}
\subsectionfont{\normalsize}
\subsubsectionfont{\bf}
\setstretch{1.125}
\setlength{\skip\footins}{0.8cm}
\setlength{\footnotesep}{0.25cm}
\setlength{\jot}{10pt}
\titlespacing*{\section}{0pt}{4pt}{4pt}
\titlespacing*{\subsection}{0pt}{15pt}{1pt}

\fancyfoot{}
\fancyfoot[LO,RE]{\vspace{-7.1pt}\includegraphics[height=9pt]{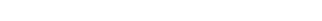}}
\fancyfoot[CO]{\vspace{-7.1pt}\hspace{11.9cm}\includegraphics{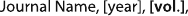}}
\fancyfoot[CE]{\vspace{-7.2pt}\hspace{-13.2cm}\includegraphics{head_foot_RF}}
\fancyfoot[RO]{\footnotesize{\sffamily{1--\pageref{LastPage} ~\textbar  \hspace{2pt}\thepage}}}
\fancyfoot[LE]{\footnotesize{\sffamily{\thepage~\textbar\hspace{4.65cm} 1--\pageref{LastPage}}}}
\fancyhead{}
\renewcommand{\headrulewidth}{0pt}
\renewcommand{\footrulewidth}{0pt}
\setlength{\arrayrulewidth}{1pt}
\setlength{\columnsep}{6.5mm}

\makeatletter
\newlength{\figrulesep}
\setlength{\figrulesep}{0.5\textfloatsep}

\newcommand{\topfigrule}{\vspace*{-1pt}%
\noindent{\color{cream}\rule[-\figrulesep]{\columnwidth}{1.5pt}} }

\newcommand{\botfigrule}{\vspace*{-2pt}%
\noindent{\color{cream}\rule[\figrulesep]{\columnwidth}{1.5pt}} }

\newcommand{\dblfigrule}{\vspace*{-1pt}%
\noindent{\color{cream}\rule[-\figrulesep]{\textwidth}{1.5pt}} }

\makeatother

\twocolumn[
  \begin{@twocolumnfalse}
{\includegraphics[height=30pt]{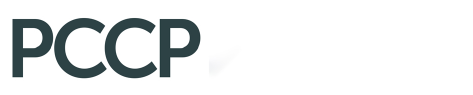}\hfill\raisebox{0pt}[0pt][0pt]{\includegraphics[height=55pt]{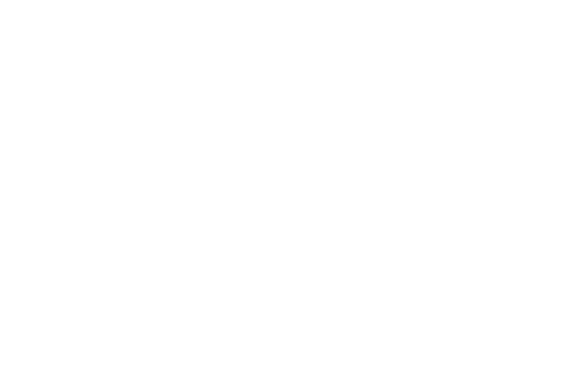}}\\[1ex]
\includegraphics[width=18.5cm]{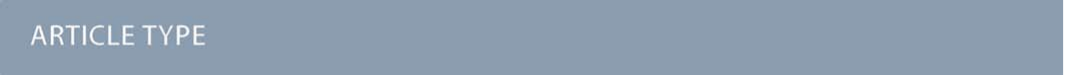}}\par
\vspace{1em}
\sffamily
\begin{tabular}{m{4.5cm} p{13.5cm} }

\includegraphics{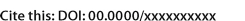} & \noindent\LARGE{\textbf{Dominant Vibronic Relaxation Channels in a Europium-based Molecular Qubit$^\dag$}} \\
\vspace{0.3cm} & \vspace{0.3cm} \\

 & \noindent\large{Neil Iyer$^{\ast}$\textit{$^{a}$}} \\

\includegraphics{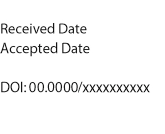} & \noindent\normalsize{%
Molecular spin qubits offer a versatile platform for quantum information processing due to their synthetic tunability and well-defined electronic structure. Here, a fitted-parameter-free computational framework combining density functional theory (DFT), time-dependent DFT (TD-DFT), and Redfield theory is applied to investigate the longitudinal spin–lattice relaxation time $T_1$ of the Eu nuclear spin qubit $\mathrm{Eu(dpphen)(NO_3)_3}$. Using a single-molecule gas-phase model, the experimental long relaxation component $T_{1,\mathrm{long}}=41.39\,\mathrm{s}$ is reproduced within a factor of 1.4 (calculated: 55.88\,s at 4.2\,K), indicating that the slow relaxation channel is governed by intramolecular vibronic coupling. In contrast, the calculated $T_{1,\mathrm{short}}$ deviates by a factor of 66, highlighting the importance of crystal lattice and intermolecular effects absent from the model. The experimental $^5D_0\to{^7F_0}$ optical transition is reproduced to within 1.1\%, supporting the accuracy of the electronic structure description. Vibrational analysis identifies a large-amplitude \textit{dpphen} rocking mode at a frequency of 332.02\,cm$^{-1}$, as the dominant vibronic coupling channel, while electric field gradient (EFG) derivative analysis independently identifies another nitrate-rocking mode at 180.57\,cm$^{-1}$ as the primary modulator of the nuclear spin environment via nitrate motion. These results are consistent with a near-maximal quadrupole asymmetry parameter $\eta=0.941$, which creates state mixing through off-diagonal quadrupolar terms. Overall, the results establish a single-molecule relaxation baseline and suggest targeted ligand rigidification and substitution strategies to suppress decoherence.} \\

\end{tabular}

 \end{@twocolumnfalse} \vspace{0.6cm}

  ]

\renewcommand*\rmdefault{bch}\normalfont\upshape
\rmfamily
\section*{}
\vspace{-1cm}

\footnotetext{\textit{$^{a}$~Independent Researcher, San Jose, CA, USA. E-mail: neil.r.iyer@gmail.com}}

\footnotetext{\dag~Electronic Supplementary Information (ESI) available. See DOI: 10.1039/cXCP00000x/}


\section{Introduction}

Quantum computing is a rapidly expanding field that uses the principles of superposition and entanglement to perform calculations that are intractable for classical computers.\cite{Raseena2025} Unlike classical bits, a qubit can exist in a linear combination of states $|0\rangle$ and $|1\rangle$.\cite{Raseena2025} Quantum computing is promising for several fields: in chemistry, quantum computers could solve electronic structure problems at polynomial cost; in cryptography, quantum algorithms are expected to dramatically accelerate decryption and factoring; and machine learning and complex optimization tasks could see significant speedups through specialized algorithms such as Grover's and Shor's.\cite{Raseena2025,Lund2025}

Molecular qubits, including those based on complex organometallic molecules as well as inorganic magnetic systems, are highly valued for their chemical tunability.\cite{Coronado2020} Unlike traditional solid-state systems, these qubits allow for atomic-precision control over the spin environment through ligand engineering.\cite{Bayliss2020} Unlike solid-state qubits, which are trapped in a specific crystal host, molecular qubits can be moved between different environments such as solutions, surfaces, or integrated into electronic and photonic devices.\cite{Bayliss2020} Additionally, because molecular qubits are synthesized through chemical assembly, they can be created in large, identical batches and organized into one-, two-, or three-dimensional configurations.\cite{Bayliss2020} Rare-earth systems in general, and $\mathrm{Eu(dpphen)(NO_3)_3}$ in particular, are promising molecular nuclear spin qubit candidates, since they have seconds-long nuclear spin lifetimes ($T_1$).\cite{Schlittenhardt2024}

Decoherence, or the loss of the stable quantum state molecular qubits rely on to function, is the key hurdle for all molecular quantum systems.\cite{Chen2020} This process is driven by several factors, most notably spin--phonon interactions, in which vibrational modes of the molecule and surrounding lattice change the spin Hamiltonian, leading to spin--lattice relaxation ($T_1$) and, more generally, decoherence processes.\cite{EscaleraMoreno2019} Nuclear spin baths may also contribute to decoherence through hyperfine interactions with nearby nuclear spins.\cite{EscaleraMoreno2019} Environmental electromagnetic field fluctuations and structural disorder can further cause decoherence and relaxation.\cite{EscaleraMoreno2019} For quantum information processing to be practical, decoherence times must be sufficiently long to enable high-fidelity gate operations, typically exceeding error-correction thresholds (usually around 99.9\%) before quantum information is lost.\cite{Barends2014}

However, a long $T_1$ also slows qubit initialization via thermal relaxation; techniques such as heat-bath algorithmic cooling can decouple initialization from $T_1$ and circumvent this limitation.\cite{Boykin2002,Baugh2005} Furthermore, the relevant figure of merit for quantum computation is $Q_M = T_2/t_{\mathrm{gate}}$, the number of coherent operations within the coherence time.\cite{Zadrozny2015} Since nuclear spin transitions are governed by the nuclear magneton $\mu_N$, which is approximately three orders of magnitude smaller than the Bohr magneton $\mu_B$, nuclear Rabi frequencies and gate speeds are correspondingly slower than for electronic spins, a trade-off that must be weighed alongside the longer $T_1$.

Building on research by Schlittenhardt \textit{et al.}\cite{Schlittenhardt2024}, which experimentally identified long nuclear spin lifetimes ($T_1$) of up to 41\,s in $\mathrm{Eu(dpphen)(NO_3)_3}$, this study aims to: (1) identify the specific intramolecular vibrations responsible for spin--phonon coupling in this complex, something only possible using computation; and (2) develop a unified model identifying ligand motions that most strongly affect both optically probed vibronic structure and ground-state nuclear spin relaxation. In this work, the qubit is defined by the nuclear spin degrees of freedom of Eu(III), while the $^5D_0 \rightarrow {}^7F_0$ optical transition is used as a spectroscopic probe and readout mechanism of the electronic structure rather than as part of the qubit manifold. Various chemical strategies to improve the already remarkable relaxation times of this complex are also proposed. 

In this work, the nuclear spin of Eu(III) defines the qubit's degrees of freedom, and the focus is on longitudinal spin--lattice relaxation ($T_1$) driven by intramolecular vibrations. The $^5D_0 \rightarrow {}^7F_0$ optical transition is not treated as part of the qubit manifold, but rather as a spectroscopic probe of the electronic structure and vibronic coupling. Accordingly, the present study focuses on phonon-mediated, non-radiative relaxation mechanisms that couple to the nuclear spin, rather than on radiative decay rates of the excited state. Radiative decay processes of the $^5D_0$ state are not explicitly modeled, as the present work focuses on vibronically mediated non-radiative channels that dominate the low-temperature relaxation dynamics considered here.

While significant progress has been made in developing first-principles approaches to describe vibronic contributions to spin relaxation,\cite{Norambuena2018,EscaleraMoreno2023} their application to specific molecular qubit systems remains an active area of research. In particular, the role of individual intramolecular vibrational modes in mediating relaxation pathways in $\mathrm{Eu(dpphen)(NO_3)_3}$ has not been explicitly identified. In this work, we employ a combination of DFT, TD-DFT, and Redfield theory to model the vibronic structure and its coupling to the spin degrees of freedom, explicitly incorporating an optically addressable excited state relevant for readout. Additionally, we include the effect of nearby nuclear spins in the molecule, which can contribute to relaxation processes, as discussed in prior studies.\cite{Norambuena2018,EscaleraMoreno2023}

\section{Methods}

All calculations were performed using DFT or TD-DFT through ORCA (version 6.0)\cite{Neese2025} and Redfield Theory through QuTiP (Quantum Toolbox in Python).\cite{Lambert2026} Coordinates of the $\mathrm{Eu(dpphen)(NO_3)_3}$ complex were obtained from the Cambridge Crystallographic Data Centre (CCDC) database.\cite{Schlittenhardt2024CCDC}

The CCDC experimental structure used for comparison was determined at 180 K, whereas the DFT-optimized geometry corresponds to a 0 K minimum on the potential energy surface.\cite{Schlittenhardt2024CCDC} This difference may introduce small deviations due to thermal motion and vibrational averaging in the experimental structure.

\subsection{DFT}

The electronic structure of the $\mathrm{Eu(dpphen)(NO_3)_3}$ molecule was determined by solving the Kohn--Sham equations\cite{Kohn1965} with DFT through the SHARK integral system\cite{Neese2022} using the \texttt{PBE} generalized gradient approximation functional.\cite{Perdew1996} DFT was used to map the ground-state potential energy surface and electronic density, which are necessary to calculate vibronic and excited state properties.

In the Unrestricted Kohn--Sham (UKS) framework, separate spatial orbitals are used for the $\alpha$ and $\beta$ spin electrons to account for the spin polarization $\xi$ of the $[\mathrm{Xe}]4f^6$ Eu(III) center, where $\xi=(\rho_\alpha(\mathbf{r}) - \rho_\beta(\mathbf{r}))/\rho(\mathbf{r})$ represents the local difference between the $\alpha$ and $\beta$ spin densities. The total energy functional $E[\rho_\alpha, \rho_\beta]$ is expressed as:

\begin{equation}
E[\rho_{\alpha}, \rho_{\beta}] = T_s[\rho_{\alpha}, \rho_{\beta}] + J[\rho_{\alpha} + \rho_{\beta}] + E_{\mathrm{ext}}[\rho_{\alpha} + \rho_{\beta}] + E_{\mathrm{xc}}[\rho_{\alpha}, \rho_{\beta}]
\end{equation}

where $T_s$ is the non-interacting kinetic energy, $J$ is the Hartree energy, $E_{\mathrm{ext}}=\int V_{\mathrm{ext}}(\mathbf{r})\rho(\mathbf{r})\,\mathrm{d}\mathbf{r}$ represents the electron--nuclear attraction, and $E_{\mathrm{xc}}$ is the exchange-correlation functional. SCF convergence was set to \texttt{TightSCF} ($10^{-8}$\,E$_\mathrm{h}$).

While density functional theory is known to have limitations in describing open-shell lanthanide systems, particularly due to strong correlation effects and the importance of spin--orbit coupling, its use here is justified by the nature of the quantities of interest. Specifically, the present study focuses on equilibrium geometries, vibrational normal modes, and electric field gradients, which depend primarily on the ground-state electron density rather than the detailed multiconfigurational character of the wavefunction. Previous studies have shown that DFT can provide reliable structural and vibrational properties for rare-earth complexes when relativistic effects are included.\cite{Neese2022} The adequacy of the electronic structure description is further assessed by comparison to experimental observables in Section~3.

\subsection{Relativistic Hamiltonian}

To properly model the heavy Eu(III) center, the Zeroth-Order Regular Approximation (ZORA) was applied to the Hamiltonian to account for relativistic effects caused by lanthanide core electrons. The electronic wavefunction was expanded in a set of atomic basis functions $\{\chi_\mu\}$, with the \texttt{SARC-ZORA-TZVP} basis set\cite{Pantazis2009} being used for Eu; the \texttt{ZORA-def2-TZVP} set\cite{Weigend2005} being used for C, N, and O; and the \texttt{ZORA-def2-SVP} set\cite{Weigend2005} being used for H.

\subsection{Hyperfine Coupling}

The magnetic interaction between the septet electronic spin ($S=3$) and the Eu nuclear spin ($I=5/2$) was evaluated through the hyperfine coupling (HFC) tensor ($\mathbf{A}$). HFC accounts for the Fermi contact interaction $A_{\mathrm{iso}}$, the magnetic dipole-dipole interaction $A_{\mathrm{dip}}$, and the spin--orbit contribution $A_{\mathrm{orb}}$ calculated using Coupled-Perturbed Kohn--Sham theory, as implemented in ORCA.\cite{Neese2003}

The hyperfine coupling tensor obtained from scalar-relativistic DFT should be interpreted with caution for Eu(III), as the true ground state is a $J=0$ multiplet arising from strong spin--orbit coupling and multiconfigurational effects.\cite{AbragamBleaney} In particular, paramagnetic spin--orbit contributions are not captured within the present framework. Consequently, the hyperfine interaction is treated here as an approximate contribution, while the nuclear quadrupole coupling, derived directly from the electric field gradient, provides the dominant relaxation mechanism.

\subsection{Nuclear Quadrupole Coupling}

For the Eu(III) nucleus, the non-spherical nuclear charge distribution interacts with the Electric Field Gradient (EFG) produced by the molecular environment, lifting the degeneracy of the nuclear spin states. This interaction is described by the Hamiltonian:\cite{Man2011}

\begin{equation}
\hat{H}_{NQC} = \frac{e^2qQ}{4I(2I-1)\hbar}\left[3\hat{I}^2_Z - I(I+1) + \eta(\hat{I}^2_X - \hat{I}^2_Y)\right]
\end{equation}

where $e^2qQ/\hbar$ is the nuclear quadrupole coupling constant; $I=5/2$ is the nuclear spin quantum number; $\eta=(V_{xx}-V_{yy})/V_{zz}$ ($0\leq\eta\leq1$) is the asymmetry parameter; and $V_{xx}$, $V_{yy}$, $V_{zz}$ are components of the EFG tensor in its principal axis frame, where $|V_{zz}|\geq|V_{yy}|\geq|V_{xx}|$.

The EFG tensor was calculated using the SHARK integral package,\cite{Neese2022} and is defined as the second derivative of the electric potential at the Eu nucleus. The EFG tensor is calculated from the total charge density $\rho(\mathbf{r})'$ as:\cite{Joosten2024}

\begin{equation}
V^{ij}_K = \int \mathrm{d}\mathbf{r'}\, \rho(\mathbf{r'})\, v^{ij}(\mathbf{r'} - \mathbf{r}_K)
\end{equation}

where $v^{ij}(\mathbf{r})=(3r^ir^j - \delta^{ij}r^2)/r^5$. The EFG derivative vibrational-coupling constants $\partial V_{\alpha\beta}/\partial Q_k$ were computed using a central-difference finite displacement approach with $\delta=0.05\,\text{Å}\,\text{amu}^{1/2}$:

\begin{equation}
\left.\frac{\partial V_{\alpha\beta}}{\partial Q_k}\right|_{Q_k=0}=
\frac{V_{\alpha\beta}(Q_k=+\delta)-V_{\alpha\beta}(Q_k=-\delta)}{2\delta}
\end{equation}

\subsection{TD-DFT}

TD-DFT was used to find the vibronic structure and Huang--Rhys factors. Excitation energies and transition densities were determined by solving the linear-response Casida equations:\cite{Casida1995}

\begin{equation}
\begin{bmatrix}
    A & B \\
    B^* & A^*
\end{bmatrix}
\begin{bmatrix}
    X \\
    Y
\end{bmatrix} = \omega
\begin{bmatrix}
    1 & 0 \\
    0 & -1
\end{bmatrix}
\begin{bmatrix}
    X \\
    Y
\end{bmatrix}
\end{equation}

where $\omega$ represents the excitation energies and $X$, $Y$ are the transition amplitudes. The first 30 roots were calculated to cover the $^5D_0\to{^7F_0}$ transition reported by Schlittenhardt \textit{et al.}\cite{Schlittenhardt2024}

\subsection{Huang--Rhys Factors}

To describe the vibrationally resolved electronic spectra, the Electronic Spectroscopy Development (ESD) module using the Vertical Gradient (VG) approximation was employed.\cite{deSouza2018} For each normal mode ($k$) with frequency ($\omega_k$), the Huang--Rhys factor ($S_k$) was determined from the displacement of the potential energy surfaces ($Q_k$) between initial and final states as $S_k = \omega_k\Delta Q_k^2/(2\hbar)$.\cite{Huang1950} The total reorganization energy $\lambda_{\mathrm{total}}=\sum S_k\hbar\omega_k$ was partitioned among the 141 vibrational degrees of freedom of the molecule.

\subsection{Redfield Theory}

To model the longitudinal relaxation ($T_1$) of the Eu nuclear spin, the Redfield Master Equation was applied, describing the dynamics of the reduced density matrix $\rho(t)$ under the Born--Markov approximation:\cite{Breuer2002,Redfield1957}

\begin{equation}
\frac{\mathrm{d}P_a(t)}{\mathrm{d}t} = \sum_b W_{ab}P_b(t)
\end{equation}

where $a$ and $b$ are eigenstates of the nuclear spin Hamiltonian, $P_a(t)$ is the probability of finding the nucleus in state $|a\rangle$ at time $t$, and $W_{ab}$ is the transition rate matrix.

The Born–Markov approximation is justified by a strong separation of timescales between nuclear spin relaxation ($T_1 \sim 10^{1}$–$10^{2}$ s) and intramolecular vibrational dynamics ($\sim 10^{-12}$–$10^{-13}$ s). Since vibrational correlation functions decay many orders of magnitude faster than spin relaxation, the bath effectively loses memory on timescales negligible compared to the spin evolution, validating a Markovian (memoryless) treatment.

\subsubsection{Computational implementation~~}
The simulation was implemented using QuTiP.\cite{Lambert2026} The total Hamiltonian $\hat{H}_{\mathrm{total}}=\hat{H}_{NQC}+\hat{H}_{HFC}$ was constructed using spin operators for $I=5/2$, with ORCA-derived values for the nuclear quadrupole coupling ($e^2qQ/\hbar=-608.79\,\mathrm{MHz}$), asymmetry ($\eta=0.941$), and isotropic hyperfine coupling ($A_{\mathrm{iso}}=-9.92\,\mathrm{MHz}$), where $\hat{H}_{NQC}$ denotes the nuclear quadrupole coupling Hamiltonian and $\hat{H}_{HFC} = \hat{\mathbf{I}} \cdot \mathbf{A} \cdot \hat{\mathbf{S}}$ denotes the hyperfine coupling Hamiltonian.

The dissipative environment was modeled using a two-channel approach without free parameters. For each vibrational mode $k$, the coupling operator $\hat{A}_k$ is defined by the projection of the EFG derivative tensor onto the Cartesian spin tensors $\hat{T}_{ij}$:

\begin{equation}
\hat{A}_k = \sum_{i,j \in \{x,y,z\}} \frac{\partial V_{ij}}{\partial Q_k} \hat{T}_{ij}
\end{equation}

The spectral density for each normal mode $Q_k$ is defined as the Fourier transform of its equilibrium autocorrelation function,\cite{Breuer2002} which for harmonic motion with phonon linewidth $\gamma_{\mathrm{ph}}$ takes the Lorentzian form:

\begin{equation}
S_k(\omega) = \frac{\hbar}{2\omega_k} \left[ (n_k + 1) \frac{\gamma_{\mathrm{ph}}/\pi}{(\omega + \omega_k)^2 + \gamma_{\mathrm{ph}}^2} + n_k \frac{\gamma_{\mathrm{ph}}/\pi}{(\omega - \omega_k)^2 + \gamma_{\mathrm{ph}}^2} \right]
\end{equation}

where $n_k = [\exp(\hbar\omega_k/k_BT)-1]^{-1}$ is the Bose--Einstein occupancy at 4.2\,K. The phonon linewidth $\gamma_{\mathrm{ph}}=20\,\mathrm{cm}^{-1}$ (HWHM) was determined from Lorentzian fits to low-frequency peaks in the DFT-computed Raman spectrum.

The $^{153}$Eu nuclear spin is coupled to 20 nearby bath spins (12 $^1$H and 8 $^{14}$N). Correlation times were estimated from the van Vleck second moment\cite{Vleck1932} of the heteronuclear bath environment, yielding $\tau_H=6.1\,\mu\mathrm{s}$ and $\tau_N=281\,\mu\mathrm{s}$. Two-phonon Raman processes and Orbach processes were excluded as negligible at 4.2\,K. The $T_1$ values were extracted using eigenmode decomposition of the transition rate matrix $W$:

\begin{equation}
h(t) = \sum_{i=1}^5 c_i\, e^{-t/T_{1,i}}
\end{equation}

where $T_{1,\mathrm{short}}$ and $T_{1,\mathrm{long}}$ correspond to the smallest and largest non-zero eigenvalues of $W$, respectively.

\section{Results}

\subsection{Geometry and electronic structure}

The DFT-optimized geometry of $\mathrm{Eu(dpphen)(NO_3)_3}$ was evaluated against the CCDC structure.\cite{Schlittenhardt2024CCDC} The coordination environment consists of the Eu(III) center with six oxygen atoms from three bidentate nitrate groups and nitrogen atoms from the nitrate groups and \textit{dpphen} ligand. Figures 1 and 2 show the molecular structure of $\mathrm{Eu(dpphen)(NO_3)_3}$, while Table~\ref{tab:bond_lengths} compares average bond lengths and angles between the optimized and CCDC geometries.

\begin{figure}[h]
\centering
\includegraphics[width=0.9\columnwidth]{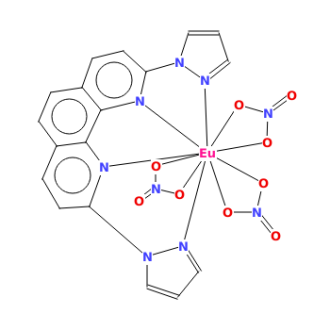}
\caption{2-D skeletal diagram of the $\mathrm{Eu(dpphen)(NO_3)_3}$ complex, obtained from the CCDC database entry.\cite{Schlittenhardt2024CCDC}}
\label{fgr:skeletal}
\end{figure}

\begin{table}[h]
\small
\caption{\ Comparison of structural bond lengths and angles between optimized and crystallographic geometries}
\label{tab:bond_lengths}
\begin{tabular*}{0.48\textwidth}{@{\extracolsep{\fill}}lccc}
\hline
\textbf{Parameter} & \textbf{ORCA (\AA\,/ \degree)} & \textbf{CIF (\AA\,/ \degree)} & $\Delta$ (\AA\,/ \degree) \\
\hline
Avg.\ Eu--O bond length & 2.5123 & 2.5030 & 0.0093 \\
Avg.\ Eu--N bond length & 2.7839 & 2.7206 & 0.0633 \\
Avg.\ N--Eu--O bond angle & 98.4 & 99.3 & 0.9 \\
Avg.\ N--Eu--N bond angle & 96.0 & 96.5 & 0.5 \\
Avg.\ O--Eu--O bond angle & 99.0 & 96.9 & 2.1 \\
\hline
\end{tabular*}
\end{table}

The optimized geometry shows exceptional agreement with experimental crystallographic data, with average percent deviations of 0.37\% for Eu--O bonds and 2.33\% for Eu--N bonds. The agreement between bond angles is also excellent across all three angle types: average deviations are $<\sim1\degree$ for N--Eu--N and N--Eu--O bonds, and $<\sim2\degree$ for O--Eu--O. These results strongly support the DFT geometry as an accurate representation of the experimental coordination sphere.

\subsection{Nuclear quadrupole coupling}

The NQC arises from the interaction of the nuclear quadrupole moment with the EFG generated by the surrounding ligands. A Quadrupole Coupling Constant of $e^2qQ/\hbar=-608.79\,\mathrm{MHz}$ and an asymmetry parameter $\eta=0.941$ were obtained. The near-maximal value of $\eta$ (approaching 1.0) is caused by the strong asymmetry of the $\mathrm{Eu(dpphen)(NO_3)_3}$ complex.

A graph of vibrational mode EFG derivatives is shown in Fig.~2. Most notably, Mode 31 ($\omega=180.57\,\mathrm{cm}^{-1}$, $S=0.004$, $\partial V_{zz}/\partial Q_k = 0.610\,\mathrm{V}\,\text{Å}^{-3}\,\mathrm{amu}^{-1/2}$) has the highest EFG derivative and is therefore the vibration to which the electronic structure is most vulnerable.

We note, however, that the magnitude of the EFG derivative alone is not sufficient to determine the contribution of a given vibrational mode to relaxation. The vibrational frequency also plays a key role through the thermal occupation factor, as higher-frequency modes are less populated at low temperatures and may therefore contribute less to relaxation despite strong coupling. In the present case, Mode 31 lies in the low-energy region of the vibrational spectrum ($\omega=180.57,\mathrm{cm}^{-1}$), where thermal population remains appreciable under typical experimental conditions. Consequently, the combination of a large EFG derivative and relatively low frequency makes this mode a particularly significant contributor to vibration-induced modulation of the electric field gradient, in line with previous analyses of how specific vibrational modes govern relaxation pathways. \cite{EscaleraMoreno2017}

To assess the stability of the central-difference EFG derivatives with respect to the displacement step, we repeated the calculation with $\delta = 0.02\,\text{Å}\,\text{amu}^{1/2}$ (versus the $\delta = 0.05$ used in the main text). The two derivative vectors element-wise to within a relative norm difference of ${\sim}5\%$, confirming numerical convergence. The resulting $T_{1,long}$ coherence time changes from ~55 to ~58~s (${\sim}5\%$), which is well within the physical uncertainties of the model.

\begin{figure}[h]
\centering
\includegraphics[width=\columnwidth]{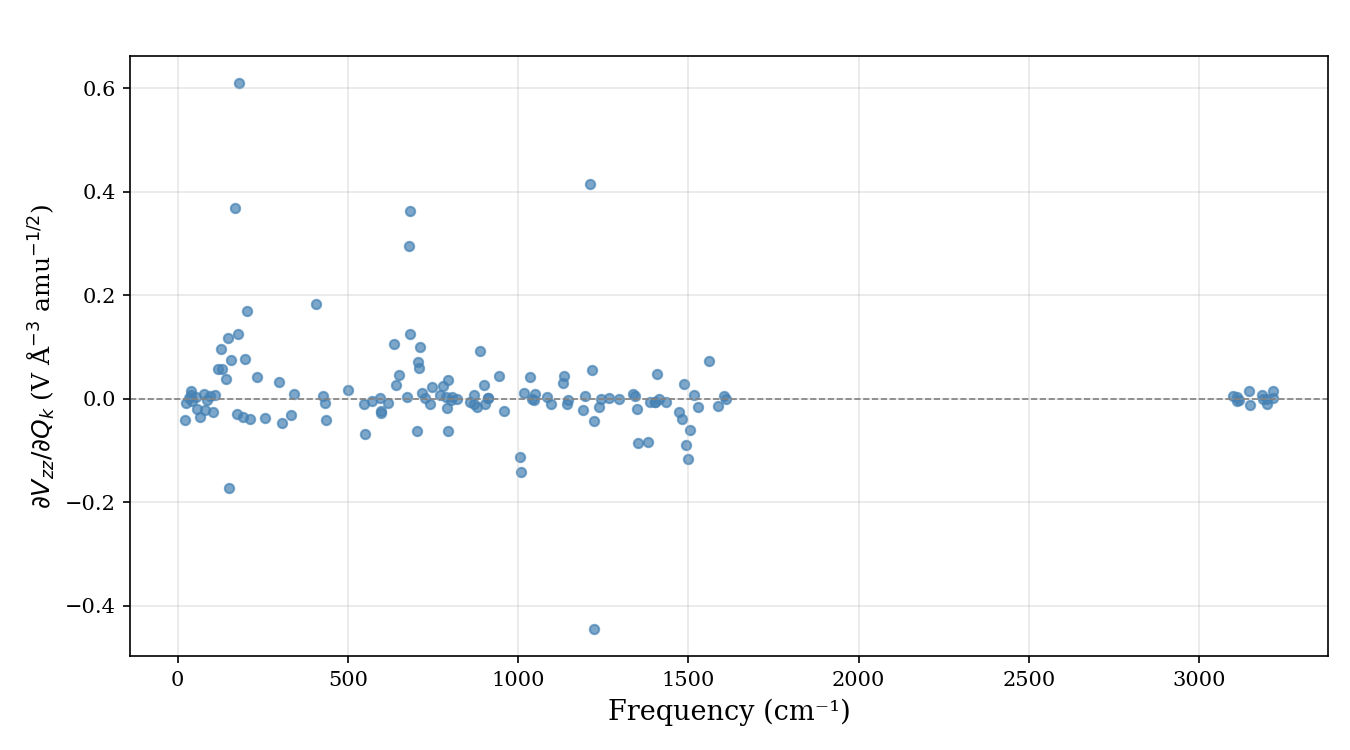}
\caption{Graph of the EFG derivative $\partial V_{zz}/\partial Q_k$ of each vibrational mode at the ground state.}
\label{fgr:efg}
\end{figure}

\subsection{Hyperfine coupling}

The HFC tensor ($\mathbf{A}$) describes the coupling between the nuclear spin and electronic spin density. In the Eu(III) septet state, the HFC is dominated by Fermi contact and spin-dipolar interactions. The calculated principal values are summarized in Table~\ref{tab:hfc_tensor}. The anisotropy confirms that the spin density is not evenly distributed around the Eu nucleus, consistent with the high asymmetry value calculated above.

\begin{table}[h]
\small
\caption{\ Principal components of the HFC tensor}
\label{tab:hfc_tensor}
\begin{tabular*}{0.48\textwidth}{@{\extracolsep{\fill}}lr}
\hline
\textbf{Component} & \textbf{Value (MHz)} \\
\hline
$A_{xx}$ & $-9.99$ \\
$A_{yy}$ & $-12.68$ \\
$A_{zz}$ & $-7.11$ \\
$A_{\mathrm{iso}}$ & $-9.92$ \\
\hline
\end{tabular*}
\end{table}

To assess the impact of the approximate hyperfine coupling treatment, we repeated the full relaxation analysis with the hyperfine interaction set to zero ($A_{iso} = 0$). The resulting $T_1$ values differ by approximately 8\% in the long relaxation component, while the short component remains essentially unchanged (the variation was <1\%). This indicates that hyperfine interactions provide only a minor correction to the relaxation dynamics, which are instead dominated by vibrationally mediated quadrupolar coupling. Consequently, the use of a scalar-relativistic DFT-based hyperfine tensor does not qualitatively affect the conclusions of this work.

The combination of NQC and HFC parameters results in an unevenly spaced energy system. The six nuclear spin states $\pm|1/2\rangle$, $\pm|3/2\rangle$, and $\pm|5/2\rangle$ are split into three quasi-degenerate pairs due to the high asymmetry ($\eta=0.941$): Pair 1 ($+|5/2\rangle$ and $+|3/2\rangle$) at $\sim$221\,MHz; Pair 2 ($\pm|1/2\rangle$) at $\sim$-35\,MHz; and Pair 3 ($-|3/2\rangle$ and $-|5/2\rangle$) at $\sim$-180\,MHz, as shown in Fig.~3.

\begin{figure}[h]
\centering
\includegraphics[width=0.8\columnwidth]{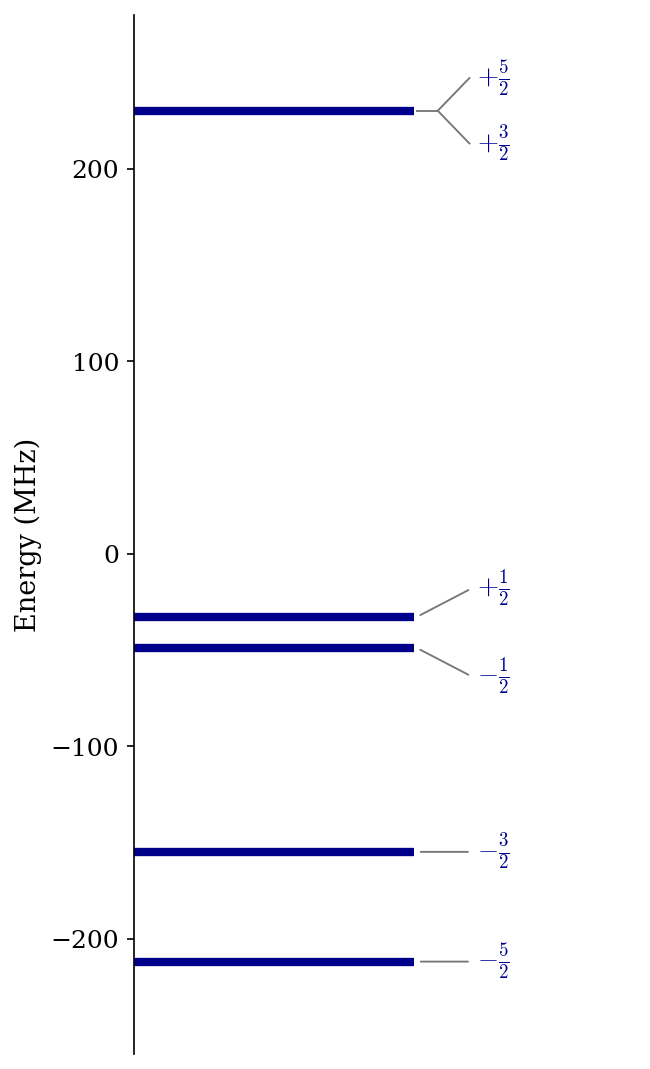}
\caption{Nuclear spin energy levels of $\mathrm{Eu(dpphen)(NO_3)_3}$.}
\label{fgr:energies}
\end{figure}

\subsection{Huang--Rhys factors}

The ESD module in ORCA identified 141 vibrational modes for $\mathrm{Eu(dpphen)(NO_3)_3}$ at the ground state. Of these, 40 modes were found to have significant coupling ($S_k>0.01$). The distribution of Huang--Rhys factors was grouped into four frequency regions: (1) low frequency $< 200\,\mathrm{cm}^{-1}$ (whole-molecule torsions); (2) mid frequency $200\text{--}1000\,\mathrm{cm}^{-1}$ (Eu--O and Eu--N stretches, highest $S_k$); (3) high frequency $1000\text{--}1600\,\mathrm{cm}^{-1}$ (internal ligand vibrations); and (4) ultra-high frequency $>3000\,\mathrm{cm}^{-1}$ (C--H stretches). The dominant vibrational modes of the ground state are listed in Table \ref{tab:hr_factors},
and all 141 vibrational modes are plotted in Fig.~4.

\begin{figure}[h]
\centering
\includegraphics[width=\columnwidth]{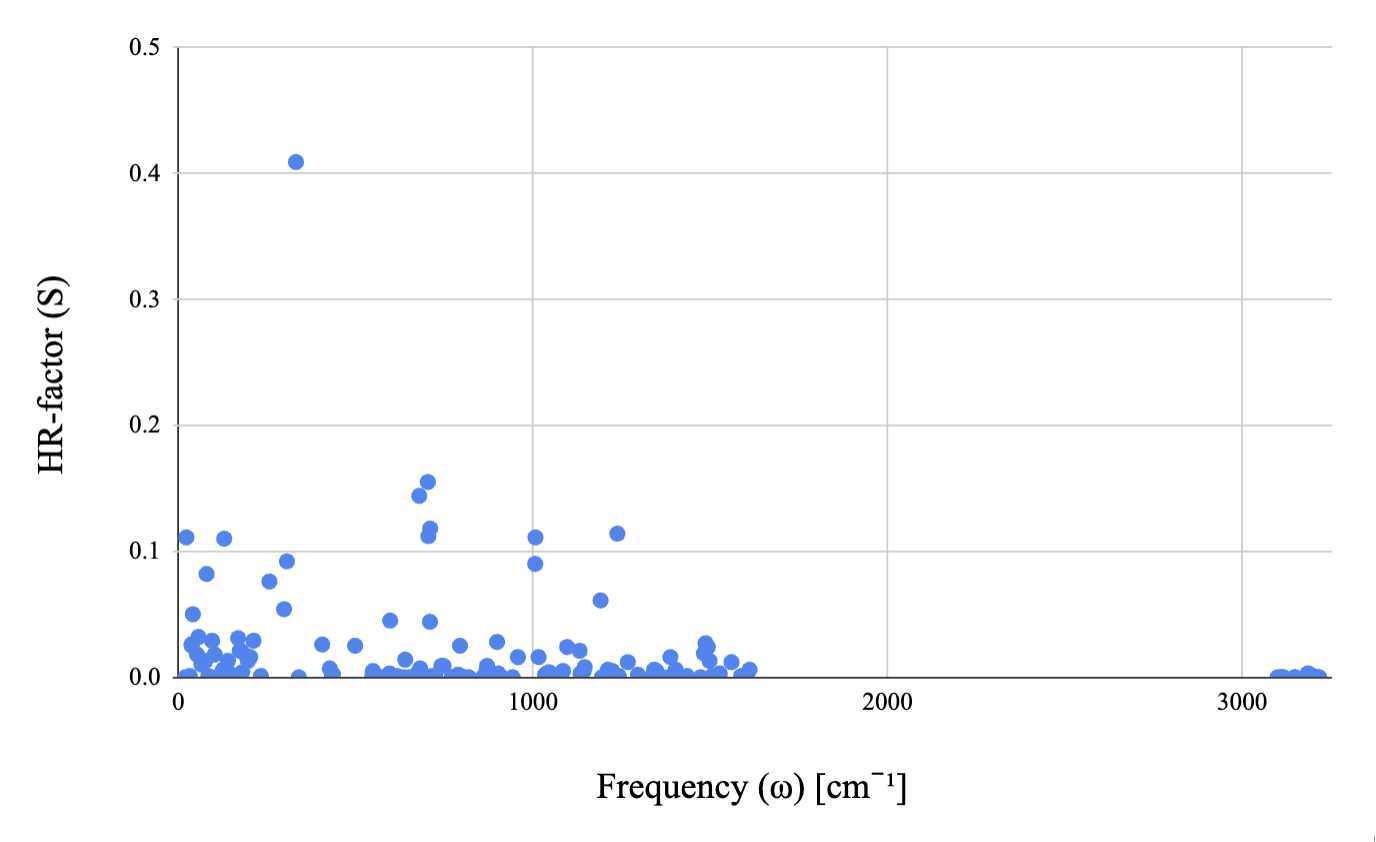}
\caption{Calculated Huang--Rhys factors for $\mathrm{Eu(dpphen)(NO_3)_3}$ at the ground state (Root 1).}
\label{fgr:hr1}
\end{figure}

\begin{table}[h]
\small
\caption{\ Dominant vibrational modes ($S_k \geq 0.1$) and Huang--Rhys factors at the ground state}
\label{tab:hr_factors}
\begin{tabular*}{0.48\textwidth}{@{\extracolsep{\fill}}lcc}
\hline
\textbf{Mode} & \textbf{Frequency (cm$^{-1}$)} & \textbf{$S_k$} \\
\hline
2   & 23.02  & 0.111 \\
19  & 129.74 & 0.110 \\
35  & 332.02 & 0.409 \\
53  & 679.79 & 0.144 \\
56  & 703.94 & 0.155 \\
57  & 705.43 & 0.112 \\
59  & 710.66 & 0.118 \\
85  & 1007.41 & 0.111 \\
103 & 1238.34 & 0.114 \\
\hline
\end{tabular*}
\end{table}

\subsection{Optical state}

The $^5D_0$ base state reported by Schlittenhardt \textit{et al.}\cite{Schlittenhardt2024} was identified as Root 24, with frequency 17,027.7\,cm$^{-1}$ (experimental: 17,212\,cm$^{-1}$) and oscillator strength $f=0.00599$, nearly ten times that of Root 25 ($\omega=17,501.2\,\mathrm{cm}^{-1}$, $f=0.00068$). The dominant vibrational modes of Root 24 are listed in Table~\ref{tab:hr_excited},  and all 141 vibrational modes of this root are plotted in Fig.~5.

\begin{figure}[h]
\centering
\includegraphics[width=\columnwidth]{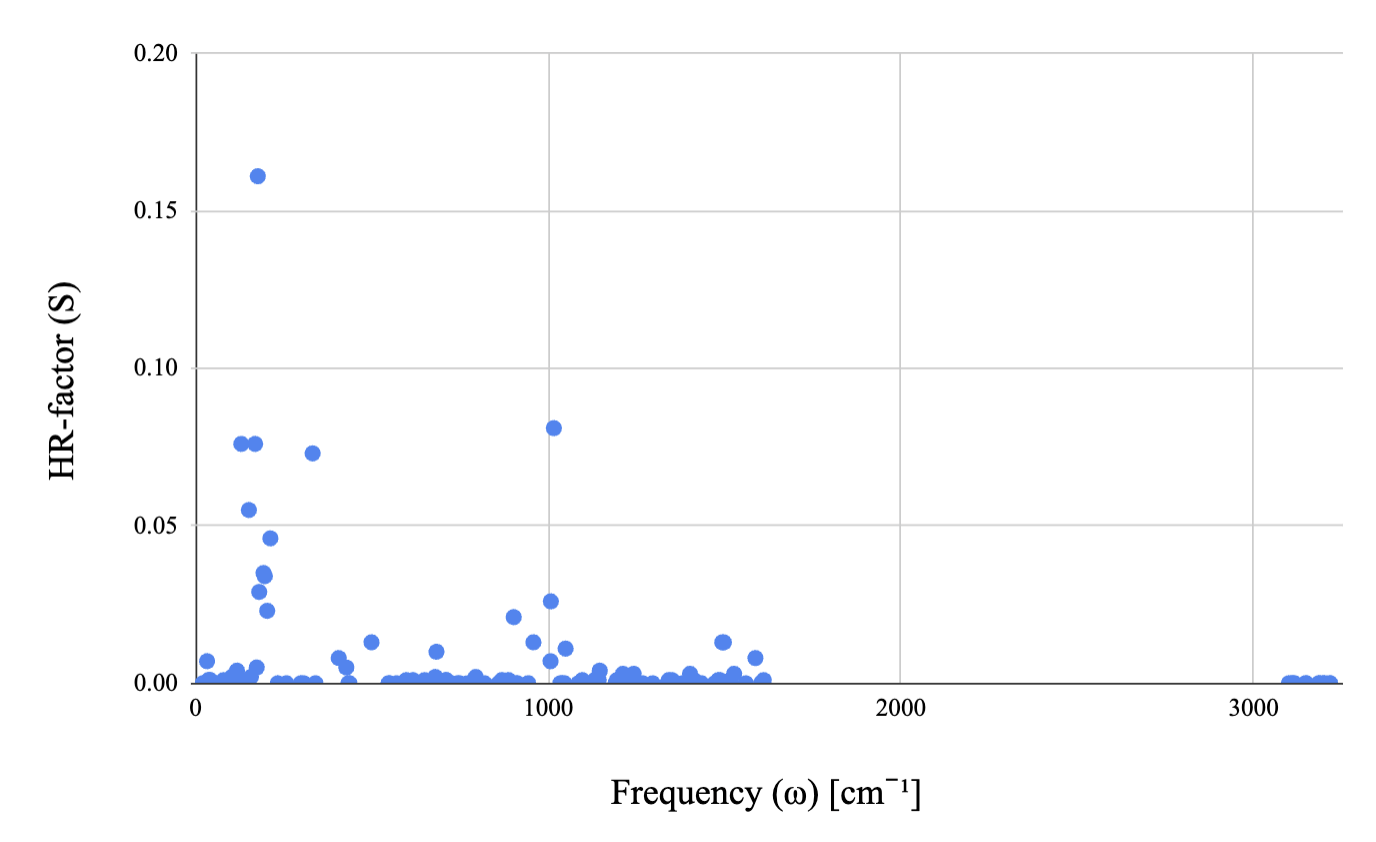}
\caption{Calculated Huang--Rhys factors for $\mathrm{Eu(dpphen)(NO_3)_3}$ at the optical excited state (Root 24).}
\label{fgr:hr2}
\end{figure}

\begin{table}[h]
\small
\caption{\ Key vibrational modes and Huang--Rhys factors for the excited state}
\label{tab:hr_excited}
\begin{tabular*}{0.48\textwidth}{@{\extracolsep{\fill}}lcc}
\hline
\textbf{Mode} & \textbf{Frequency (cm$^{-1}$)} & \textbf{$S_k$} \\
\hline
18 & 129.74 & 0.076 \\
21 & 151.25 & 0.055 \\
23 & 168.89 & 0.076 \\
25 & 176.65 & 0.161 \\
35 & 332.02 & 0.073 \\
86 & 1015.95 & 0.081 \\
\hline
\end{tabular*}
\end{table}

\subsection{Redfield theory and $T_1$}

The simulation yielded $T_{1,\mathrm{long}}=55.88\,\mathrm{s}$ (experimental: 41.39\,s)\cite{Schlittenhardt2024} and $T_{1,\mathrm{short}}=20.35\,\mathrm{s}$ (experimental: 0.31\,s).\cite{Schlittenhardt2024} The long result is 1.34 times the experimental value, while the short coherence time is 65.6 times off the experimental value. The simulated and experimental population decays are shown in Fig.~6.

\begin{figure}[h]
\centering
\includegraphics[width=\columnwidth]{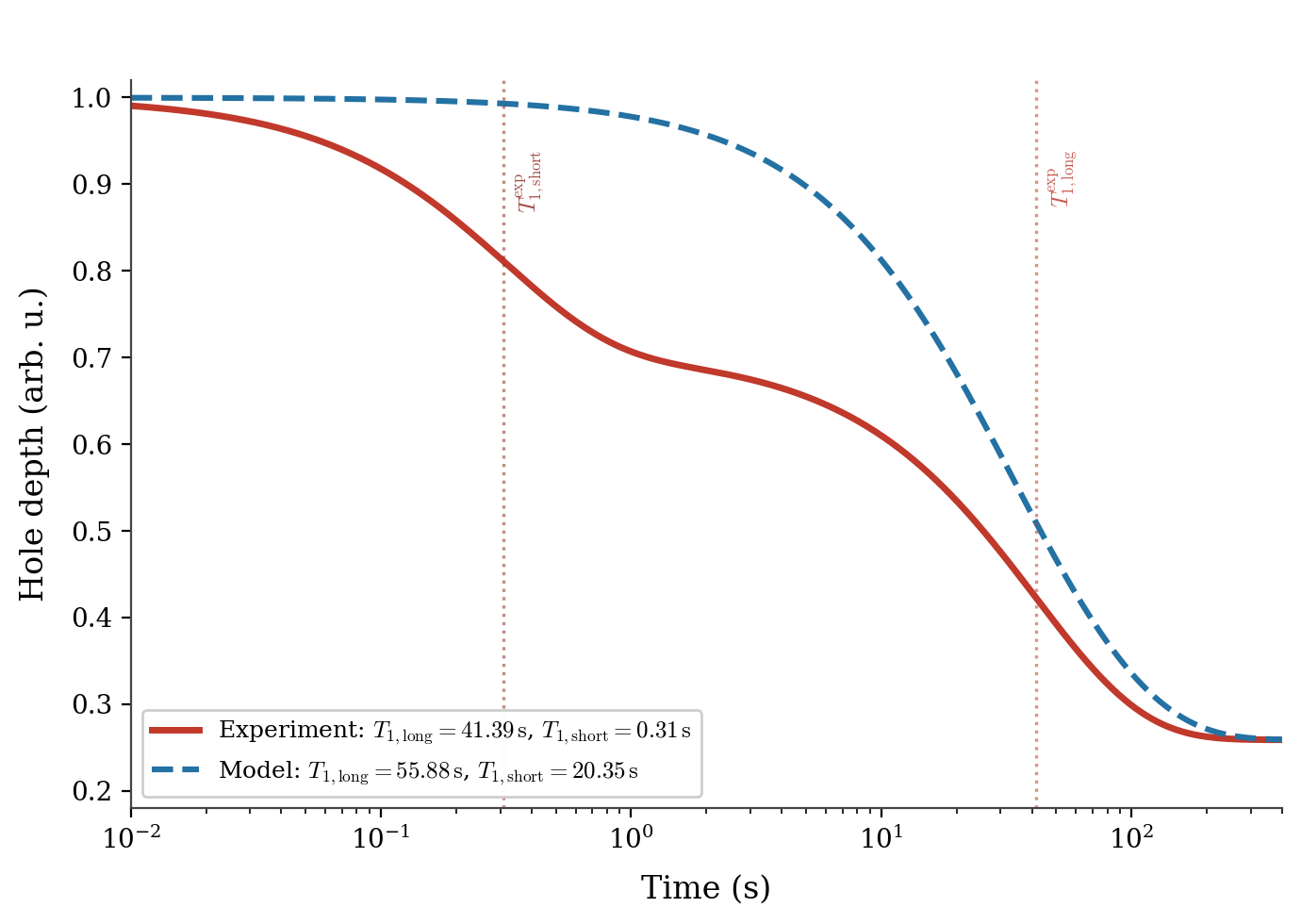}
\caption{$T_1$ population decay of $\mathrm{Eu(dpphen)(NO_3)_3}$. Blue dashed: calculated; red-orange solid: experimental. Hole depth (a.u.) represents the longitudinal magnetization.}
\label{fgr:t1}
\end{figure}

Sensitivity analysis confirms that relaxation eigenvalues were invariant for secular cutoffs spanning $0.001\text{--}10\,\mathrm{rad\,s}^{-1}$. Varying $\gamma_{\mathrm{ph}}$ over $5\text{--}50\,\mathrm{cm}^{-1}$ altered absolute $T_1$ values by approximately a factor of three but preserved a nearly constant ratio $T_{1,\mathrm{long}}:T_{1,\mathrm{short}} \approx 2.3\text{--}3.7$, far below the experimental 134$\times$ separation. Artificial introduction of low-frequency phonon modes (0.5--2\,cm$^{-1}$) uniformly accelerated both relaxation channels while preserving an approximately constant ratio, indicating that missing low-frequency intramolecular modes are insufficient to account for the experimentally observed fast component.

\section{Discussion}

\subsection{Structural influences on relaxation}

A key finding is the near-maximal asymmetry parameter $\eta=0.941$. This extreme asymmetry, caused by the low-symmetry coordination environment of the \textit{dpphen} and nitrate ligands, leads to an unevenly spaced energy level scheme that allows for the mixing of energy states. Formally, a nonzero $\eta$ introduces off-diagonal terms $\hat{H}_{\mathrm{NQC}} \propto \eta(I_+^2+I_-^2)$ that mix nuclear spin states and allow normally forbidden transitions to occur, decreasing coherence times.

The observation of such a high asymmetry parameter raises the question of whether the molecule should be modified to create a more idealized, high-symmetry coordination environment. While this would theoretically reduce state mixing and decoherence, in a highly symmetric environment the $^5D_0\to{^7F_0}$ transition could become strictly forbidden, preventing optical readout. There is therefore a genuine design tradeoff: minimizing quadrupolar state mixing favors $\eta\to0$, while preserving optical addressability requires $\eta\neq0$. Rather than eliminating the asymmetry, the more productive design strategy is to reduce vibronic coupling, which is the dominant driver of decoherence at non-cryogenic temperatures.

The 141 vibrational modes of the ground-state geometry were analyzed through two complementary lenses: the Huang--Rhys factor $S_k$, which quantifies the displacement of the nuclear geometry along a given normal coordinate between electronic states; and the EFG derivative $\partial V_{zz}/\partial Q_k$, which quantifies the sensitivity of the principal EFG component to motion along each normal coordinate.

For optical decoherence, Mode 35 at 332.02\,cm$^{-1}$ is the dominant contributor in the electronic ground state, with $S=0.409$. This mode is characterized by motion localized on the \textit{dpphen} ligand, with pyrazolyl ring displacements of up to 0.06\,\AA\ and phenanthroline displacements of up to 0.04\,\AA, while Eu--O bond distances and the Eu center remain essentially stationary. At 4.2\,K, thermal population of this mode is negligible; however, zero-point fluctuations persist at all temperatures, corresponding to \textit{dpphen} displacements of approximately 0.03--0.04\,\AA. At room temperature, the Bose--Einstein occupation reaches $n\approx0.2$, amplifying effective nuclear displacements by a factor of $\sqrt{2n+1}\approx1.2$. Mode 35 remains a significant contributor in the excited optical state ($S=0.073$), though in this manifold it is superseded by Mode 25 at 176.65\,cm$^{-1}$ ($S=0.161$), which has a substantially higher thermal occupation ($n\approx0.7$ at 300\,K) and is the primary driver of optical decoherence at room temperature.

For nuclear spin decoherence, EFG derivative analysis identifies Mode 31 at 180.57\,cm$^{-1}$ ($\partial V_{zz}/\partial Q_k = 0.610\,\mathrm{V\,\text{\AA}^{-3}\,amu^{-1/2}}$, $S=0.004$) as the dominant modulator of the nuclear spin environment, a nitrate oxygen rocking mode entirely distinct from the \textit{dpphen} modes governing optical decoherence.

These results suggest two primary structural engineering targets: (1) stiffening the \textit{dpphen} ligand and (2) replacing the nitrate ligands with more geometrically constraining alternatives.

\subsection{Causes of the $T_1$ discrepancy}

The 65.7$\times$ discrepancy between the calculated (20.35\,s) and experimental\cite{Schlittenhardt2024} $T_{1,\mathrm{short}}$ (0.31\,s) is primarily attributed to the limitations of the single-molecule model. The crystal used in Schlittenhardt \textit{et al.}'s experiment likely introduces additional relaxation mechanisms:

(1) \textit{Collective lattice phonons}: While artificial introduction of low-frequency modes within the isolated-molecule framework uniformly accelerated both relaxation components without generating the experimentally observed 134$\times$ separation, collective lattice phonons that modulate intermolecular distances or mediate long-range dipolar coupling may introduce distinct relaxation pathways absent in the single-molecule model.

(2) \textit{Intermolecular dipolar coupling}: In a crystalline environment, Eu centres and surrounding nuclei form an extended dipolar network. Such long-range interactions can create additional relaxation eigenmodes and enable cross-relaxation pathways not captured in a localized bath treatment, providing a plausible origin of the experimentally observed $T_{1,\mathrm{short}}$.

(3) \textit{Crystal imperfections}: Lattice defects, impurities, and surface states can introduce electromagnetic noise that accelerates decoherence beyond what is captured in a single-molecule model.

Additionally, the present DFT-based treatment does not include the paramagnetic spin--orbit (PSO) contribution to the hyperfine interaction, which can be significant in lanthanide systems. This omission represents a known limitation of single-reference approaches for heavy elements and is not expected to qualitatively affect the identification of dominant vibrational relaxation channels.

Despite these omissions, the present results demonstrate that intramolecular vibronic coupling sets an intrinsic baseline for nuclear spin relaxation, accurately reproducing the slow relaxation component. The failure of extensive low-frequency perturbations to generate the fast channel within the isolated-molecule framework indicates that this component likely originates from collective solid-state mechanisms.

\section*{Conclusions}

A computational framework combining DFT, TD-DFT, and Redfield Theory was employed to investigate the longitudinal relaxation ($T_1$) of the rare-earth nuclear spin molecular qubit $\mathrm{Eu(dpphen)(NO_3)_3}$. The simulation predicted $T_{1,\mathrm{long}}=55.88\,\mathrm{s}$, of the same order of magnitude as the experimental result of Schlittenhardt \textit{et al.}\cite{Schlittenhardt2024} ($T_{1,\mathrm{long}}=41.39\,\mathrm{s}$), confirming that a simple and inexpensive computational approach can accurately model the relaxation of a molecular qubit at cryogenic temperatures. However, the model did not account for crystal lattice imperfections and intermolecular coupling, leading to a $T_{1,\mathrm{short}}$ value 65.7 times off the experimental one.

Two main factors govern the coherence of the $\mathrm{Eu(dpphen)(NO_3)_3}$ system. First, the high structural asymmetry $\eta=0.941$ leads to significant state mixing, allowing normally forbidden relaxation pathways to occur. Second, Mode 35 (332.02\,cm$^{-1}$) is characterized by large-scale rocking of the \textit{dpphen} ligand, with Huang--Rhys factors of 0.409 (ground state) and 0.073 (optical state), making it the primary channel for optical decoherence. Complementary EFG derivative analysis identifies Mode 31 at 180.57\,cm$^{-1}$ as the dominant modulator of the nuclear spin environment ($\partial V_{zz}/\partial Q_k = 0.610\,\mathrm{V\,\text{\AA}^{-3}\,amu^{-1/2}}$).

Two distinct chemical engineering strategies are proposed to mitigate vibronic decoherence. For the \textit{dpphen} scaffold, we propose: (1) \textit{Steric locking} --- introduction of bulky substituents such as \textit{tert}-butyl or adamantyl groups to increase the zero-point energy of Mode 35; and (2) \textit{Structural rigidification} --- replacing the C--N bonds connecting phenanthroline to each pyrazolyl group with fused aromatic rings. For the nitrate co-ligands, we propose the replacement of the three $\mathrm{NO_3^-}$ groups with more rigid aromatic chelators such as oxalate, fluorinated dicarboxylate, or acetylacetonate rings, which enforce constrained O-donor bite angles through rigid chelate rings and suppress Eu--O rocking motion.

In an extension of this work, periodic DFT methods (DFPT) within a periodic supercell approach will be employed to compute phonon dispersion relations and the modulation of EFG tensors by lattice vibrations, addressing the discrepancy in $T_{1,\mathrm{short}}$ and enabling systematic evaluation of proposed chemical modifications. We will also test the stability of this molecular qubit at temperatures above 4.2 K.

\section*{Data availability}

All electronic structure calculations were performed using the ORCA 6.0 program system\cite{Neese2025} and the SHARK integral package.\cite{Neese2022} Software is freely available at \url{https://www.faccts.de/orca/}. Redfield Theory calculations were performed using QuTiP,\cite{Lambert2026} available at \url{https://github.com/qutip/qutip}. Optimized geometry files (.xyz), ORCA input/output files (.inp/.out), and the Python relaxation simulation script (.py) are available at \url{https://github.com/neili7/ORCA-and-Qutip-Calculations}.

\section*{Author contributions}
N.\ Iyer: conceptualization, methodology, investigation, formal analysis, writing -- original draft.

\section*{Conflicts of interest}
There are no conflicts to declare.

\section*{Acknowledgements}

The author expresses sincere thanks to Professor Brenda Rubenstein of Brown University for invaluable mentorship throughout this research. Thanks are also due to the Brown University Center for Computation and Visualization (CCV) for access to the OSCAR high-performance computing cluster.

\balance

\bibliographystyle{rsc}
\bibliography{references}

\end{document}